\RequirePackage{lineno}
\documentclass[final,5p,times,twocolumn]{elsarticle}
\include{threephase_definitions}

\usepackage{color}      
\usepackage{graphicx}
\usepackage{subfigure}
\usepackage{verbatim}
\usepackage{multirow}
\usepackage{epstopdf}

\newcommand{\figuresize}{0.95\columnwidth}

\journal{Nuclear Physics A}

\begin{document}

\begin{frontmatter}

\title{Calibration of Muon Reconstruction Algorithms Using an External Muon Tracking System at the Sudbury Neutrino Observatory}

\def\altaffiliation#1{\footnote{#1}}
\newcommand{\alta}{Department of Physics, University of 
Alberta, Edmonton, Alberta, T6G 2R3, Canada}
\newcommand{\ubc}{Department of Physics and Astronomy, University of 
British Columbia, Vancouver, BC V6T 1Z1, Canada}
\newcommand{\bnl}{Chemistry Department, Brookhaven National 
Laboratory,  Upton, NY 11973-5000}
\newcommand{\carleton}{Ottawa-Carleton Institute for Physics, Department of Physics, Carleton University, Ottawa, Ontario K1S 5B6, Canada}
\newcommand{\uog}{Physics Department, University of Guelph,  
Guelph, Ontario N1G 2W1, Canada}
\newcommand{\lu}{Department of Physics and Astronomy, Laurentian 
University, Sudbury, Ontario P3E 2C6, Canada}
\newcommand{\lbnl}{Institute for Nuclear and Particle Astrophysics and 
Nuclear Science Division, Lawrence Berkeley National Laboratory, Berkeley, CA 94720}
\newcommand{\lbla}{ Lawrence Berkeley National Laboratory, Berkeley, CA}
\newcommand{\lanl}{Los Alamos National Laboratory, Los Alamos, NM 87545}
\newcommand{\llnl}{Lawrence Livermore National Laboratory, Livermore, CA}
\newcommand{\lanla}{Los Alamos National Laboratory, Los Alamos, NM 87545}
\newcommand{\oxford}{Department of Physics, University of Oxford, 
Denys Wilkinson Building, Keble Road, Oxford OX1 3RH, UK}
\newcommand{\penn}{Department of Physics and Astronomy, University of 
Pennsylvania, Philadelphia, PA 19104-6396}
\newcommand{\queens}{Department of Physics, Queen's University, 
Kingston, Ontario K7L 3N6, Canada}
\newcommand{\uw}{Center for Experimental Nuclear Physics and Astrophysics, 
and Department of Physics, University of Washington, Seattle, WA 98195}
\newcommand{\uta}{Department of Physics, University of Texas at Austin, Austin, TX 78712-0264}
\newcommand{\triumf}{TRIUMF, 4004 Wesbrook Mall, Vancouver, BC V6T 2A3, Canada}
\newcommand{\ralimp}{Rutherford Appleton Laboratory, Chilton, Didcot OX11 0QX, UK}
\newcommand{\iusb}{Department of Physics and Astronomy, Indiana University, South Bend, IN}
\newcommand{\fnal}{Fermilab, Batavia, IL}
\newcommand{\uo}{Department of Physics and Astronomy, University of Oregon, Eugene, OR}
\newcommand{\hu}{School of Engineering, Hiroshima University, Hiroshima, Japan}
\newcommand{\slac}{Stanford Linear Accelerator Center, Menlo Park, CA}
\newcommand{\mac}{Department of Physics, McMaster University, Hamilton, ON}
\newcommand{\doe}{US Department of Energy, Germantown, MD}
\newcommand{\lund}{Department of Physics, Lund University, Lund, Sweden}
\newcommand{\mpi}{Max-Planck-Institut for Nuclear Physics, Heidelberg, Germany}
\newcommand{\uom}{Ren\'{e} J.A. L\'{e}vesque Laboratory, Universit\'{e} de Montr\'{e}al, Montreal, PQ}
\newcommand{\cwru}{Department of Physics, Case Western Reserve University, Cleveland, OH}
\newcommand{\pnnl}{Pacific Northwest National Laboratory, Richland, WA}
\newcommand{\uc}{Department of Physics, University of Chicago, Chicago, IL}
\newcommand{\mitt}{Laboratory for Nuclear Science, Massachusetts Institute of Technology, Cambridge, MA 02139}
\newcommand{\ucsd}{Department of Physics, University of California at San Diego, La Jolla, CA }
\newcommand{	\lsu	}{Department of Physics and Astronomy, Louisiana State University, Baton Rouge, LA 70803}
\newcommand{\imp}{Imperial College, London SW7 2AZ, UK}
\newcommand{\uci}{Department of Physics, University of California, Irvine, CA 92717}
\newcommand{\ucia}{Department of Physics, University of California, Irvine, CA}
\newcommand{\suss}{Department of Physics and Astronomy, University of Sussex, Brighton  BN1 9QH, UK}
\newcommand{	\lifep	}{Laborat\'{o}rio de Instrumenta\c{c}\~{a}o e F\'{\i}sica Experimental de
Part\'{\i}culas, Av. Elias Garcia 14, 1$^{\circ}$, 1000-149 Lisboa, Portugal}
\newcommand{\hku}{Department of Physics, The University of Hong Kong, Hong Kong.}
\newcommand{\aecl}{Atomic Energy of Canada, Limited, Chalk River Laboratories, Chalk River, ON K0J 1J0, Canada}
\newcommand{\nrc}{National Research Council of Canada, Ottawa, ON K1A 0R6, Canada}
\newcommand{\princeton}{Department of Physics, Princeton University, Princeton, NJ 08544}
\newcommand{\birkbeck}{Birkbeck College, University of London, Malet Road, London WC1E 7HX, UK}
\newcommand{\snoi}{SNOLAB, Sudbury, ON P3Y 1M3, Canada}
\newcommand{\uba}{University of Buenas Aires, Argentina}
\newcommand{\hvd}{Department of Physics, Harvard University, Cambridge, MA}
\newcommand{\pny}{Goldman Sachs, 85 Broad Street, New York, NY}
\newcommand{\pnv}{Remote Sensing Lab, PO Box 98521, Las Vegas, NV 89193}
\newcommand{\psis}{Paul Schiffer Institute, Villigen, Switzerland}
\newcommand{\liverpool}{Department of Physics, University of Liverpool, Liverpool, UK}
\newcommand{\uto}{Department of Physics, University of Toronto, Toronto, ON, Canada}
\newcommand{\uwisc}{Department of Physics, University of Wisconsin, Madison, WI}
\newcommand{\psu}{Department of Physics, Pennsylvania State University,
     University Park, PA}
\newcommand{\anl}{Deparment of Mathematics and Computer Science, Argonne
     National Laboratory, Lemont, IL}
\newcommand{\cornell}{Department of Physics, Cornell University, Ithaca, NY}
\newcommand{\tufts}{Department of Physics and Astronomy, Tufts University, Medford, MA}
\newcommand{\ucd}{Department of Chemical Engineering and Materials Science, University of California, Davis, CA}
\newcommand{\unc}{Department of Physics, University of North Carolina, Chapel Hill, NC}
\newcommand{\dresden}{Institut f\"{u}r Kern- und Teilchenphysik, Technische Universit\"{a}t Dresden,  01069 Dresden, Germany}
\newcommand{\fargo}{Business Direct, Wells Fargo, San Francisco, CA}
\newcommand{\ucol}{Physics Department, University of Colorado at Boulder, Boulder, CO}
\newcommand{\utah}{University of Utah Department of Physics, Salt Lake City, Utah}
\newcommand{\ucsb}{Department of Physics, University of California Santa Barbara, Santa Barbara, CA }
\newcommand{\cern}{CERN (European Laboratory for Particle Physics), Geneva, Switzerland}
\newcommand{\ISS}{Institute for Space Sciences, Freie Universit\"{a}t, Berlin, Leibniz-Institute of Freshwater Ecology and Inland Fisheries, Germany}


\author[mitt]{R.~Abruzzio}
\author[lbnl]{Y.D.~Chan}
\author[lbnl]{C.A.~Currat\altaffiliation{Current Address: \fargo}}
\author[snoi,queens]{F.A.~Duncan}
\author[lu]{J.~Farine}
\author[snoi]{R.J.~Ford}
\author[mitt,uw]{J.A.~Formaggio\corref{cor1}}
\ead{josephf@mit.edu}
\author[uw,queens,lbnl,lanl]{N.~Gagnon}
\author[alta,queens]{A.L.~Hallin}
\author[queens,lanl,ubc]{J.~Heise}
\author[unc,uw]{M.A.~Howe}
\author[mitt]{E.~Ilhoff}
\author[mitt]{J.~Kelsey}
\author[penn]{J.R.~Klein}
\author[alta,queens]{C.~Kraus}
\author[lu]{A.~Kr\"{u}ger}
\author[lsu]{T.~Kutter}
\author[penn]{C.C.M.~Kyba\altaffiliation{Current address: \ISS}}
\author[snoi,uog]{I.T.~Lawson}
\author[lbnl]{K.T.~Lesko}
\author[penn]{N.~McCauley\altaffiliation{Current address: \liverpool}}
\author[mitt]{B.~Monreal \altaffiliation{Current address: \ucsb}}
\author[mitt]{J.~Monroe}
\author[queens]{A.J.~Noble}
\author[mitt]{R.A.~Ott}
\author[lbnl]{A.W.P.~Poon}
\author[lbnl]{G.~Prior\altaffiliation{Current address: \cern}}
\author[lanl,uw]{K.~Rielage}
\author[mitt,queens]{T.J.~Sonley}
\author[ubc]{T.~Tsui}
\author[uw]{B.~Wall}
\author[unc,uw]{J.F.~Wilkerson}

\address[mitt]{\mitt}  
\address[lbnl]{\lbnl} 
\address[snoi]{\snoi}  
\address[queens]{\queens}  
\address[lu]{\lu}     
\address[uw]{\uw}  
\address[lanl]{\lanl} 
\address[alta]{\alta}  
\address[ubc]{\ubc}   
\address[unc]{\unc} 
\address[penn]{\penn} 
\address[lsu]{\lsu}   
\address[uog]{\uog} 

\cortext[cor1]{Corresponding author.}

\date{\today}

\begin{abstract}
To help constrain the algorithms used in reconstructing high-energy muon events incident on the Sudbury Neutrino Observatory (SNO), a muon tracking system was installed.  The system consisted of four planes of wire chambers, which were triggered by scintillator panels. The system was integrated with SNO's main data acquisition system and took data for a total of 95 live days.  Using cosmic-ray events reconstructed in both the wire chambers and in SNO's water Cherenkov detector, the external muon tracking system was able to constrain the uncertainty on the muon direction to better than 0.6$^\circ$.
\end{abstract}
\end{frontmatter}


\section{Introduction}
\label{sec:intro}

The Sudbury Neutrino Observatory (SNO) was a large water Cherenkov detector optimized for detecting solar neutrinos created from the $^8$B reaction in the main $pp$ fusion chain. In addition to solar neutrinos, the Sudbury Neutrino Observatory was also sensitive to high-energy muons that traverse the volume of the detector.  A small fraction of these events are neutrino-induced muons from atmospheric neutrinos, while the large remaining fraction come from cosmic rays created in the upper atmosphere.  It is possible to discriminate between these muon sources by looking at the angular distribution of incoming muons.  The combination of large depth and the relatively flat topography in the vicinity of the detector attenuates almost all cosmic ray muons entering the detector at zenith angle $\cos{(\theta_z)} > 0.4$.  The study of muon events in the SNO detector provides measurements of the absolute flux of atmospheric neutrinos and constraints on the atmospheric neutrino mixing parameters $\Delta m^2_{23}$ and $\theta_{23}$~\cite{bib:SNO_atmo}.  While the latter measurement is more strongly constrained by other experiments~\cite{bib:SK, bib:Minos}, the former is unique to the SNO experiment.

To facilitate a clean measurement of the zenith distribution of muons entering the SNO fiducial volume, an accurate understanding of the muon reconstruction algorithm is necessary.  This includes both the angular and spatial resolution of high-energy muons which enter the detector.  Determining the accuracy of the muon tracking reconstruction algorithm, however, relies almost entirely on Monte Carlo simulations.  Although the detector response to muons was benchmarked against selected cosmic-ray data, there is not an external calibration source that can provide a consistency-check to the accuracy of the reconstruction algorithm.  This is in sharp contrast to the case for SNO's response to neutrons and low energy electrons, which was calibrated with multiple sources to a precision of $\sim 1\%$~\cite{bib:sno_cal}.

We present in this paper a means by which the SNO experiment was able to calibrate its muon tracking algorithm via the use of an external muon tracking system.  The External Muon System (EMuS) allowed SNO to simultaneously reconstruct selected cosmic-ray events in two independent systems, thereby providing a cross-check on the tracking algorithm.  The EMuS experiment ran for a total of 94.6 live days during the last phase of the SNO experiment. 

This paper is divided as follows: Section \ref{sec:SNO} describes the main SNO experiment, Section~\ref{sec:FTI} describes the SNO muon reconstruction algorithm, Section~\ref{sec:EMuS} describes the characteristics of the EMuS apparatus, Section~\ref{sec:data_selection} describes the criterion for accepting events, and finally Section~\ref{sec:recon} discusses the analysis used to calibrate the SNO tracking algorithm against data taken with the EMuS system.

\section{The Sudbury Neutrino Observatory}
\label{sec:SNO}

The SNO detector consisted of a 12-meter-diameter acrylic sphere filled with 1 kiloton of D$_2$O. The 5.5-cm-thick acrylic vessel was surrounded by 7.4 kilotons of ultra-pure H$_2$O encased within a barrel-shaped cavity, 34 m in height and 22 m in diameter. A 17.8-meter-diameter geodesic structure surrounded the acrylic vessel and supported 9456 20-cm-diameter photomultiplier tubes (PMTs) pointed toward the center of the detector. A non-imaging light concentrator was mounted on each PMT to increase the effective photocathode coverage to 54\%~\cite{bib:Doucas}. The detector is described in detail elsewhere~\cite{bib:sno_nim}. 

SNO was located in the Vale Creighton mine in Ontario, Canada at a depth of 2.092 km ($5890 \pm 94$ meters water equivalent) with a flat overburden. At this depth, the muon rate incident over the geodesic sphere and integrated over the seasonal variation is $62.9 \pm 0.2~ \mu$/day across an impact area of 216 m$^2$\cite{bib:SNO_atmo}. Muons entering the detector produce Cherenkov light at an angle of 42$^\circ$ with respect to the propagation direction of the muon. Cherenkov light and light from delta rays produced by the muon illuminate an average of 5500 PMTs, whose charge and timing information are recorded. The amplitude and timing response of the PMTs were calibrated {\it in situ} using a light diffusing sphere illuminated by a laser at six distinct wavelengths \cite{bib:sno_cal}. This laser ball calibration was of particular relevance to the muon fitter because it provides a timing and charge calibration for multiple photon hits on a single PMT. Other calibration sources used in SNO are described elsewhere \cite{bib:sno_nim, bib:sno_cal2}.

Data taking in the SNO experiment was subdivided into three distinct phases for measurement of the solar neutrino flux. In the first phase, the experiment ran with pure D$_2$O only. The solar neutral current reaction was observed by detecting the 6.25 MeV $\gamma$-ray following the capture of the neutron by a deuteron. For the second phase of data taking, approximately 0.2\% by weight of purified NaCl was added to the D$_2$O volume to enhance the sensitivity to neutrons via their capture on $^{35}$Cl. In the third and final phase of the experiment, 40 discrete $^3$He and $^4$He proportional tubes were inserted within the fiducial volume of the detector.  This enhanced the neutron capture cross-section to make an independent measurement of the neutron flux, by observing neutron capture on $^3$He in the proportional counters. Results from the measurements of the solar neutrino flux for these phases have been reported elsewhere~\cite{bib:sno1,bib:sno2,bib:sno3,bib:sno4,bib:sno5,bib:sno6, bib:sno7}.

\section{Muon Reconstruction with the SNO Detector}
\label{sec:FTI}

The SNO muon reconstruction algorithm fits for a through-going muon track based on the charge, timing, and spatial distribution of triggered PMTs. Using a maximum likelihood method, the fitter is able to determine a variety of muon tracking parameters, including the muon's propagation direction, impact parameter with respect to the center of SNO, the total deposited energy, and a timing offset. The likelihood is defined as:
\begin{equation}
L = \prod_i^{PMTs} \left[ \sum^{\infty}_{n=1} P_N(n|\lambda_i) P_Q(Q_i|n) P_T(t_i|n) \right]
\label{eq:FTI_lh}
\end{equation}
where $n$ is the number of detected photons, $P_N(n|\lambda_i)$ is the probability of $n$ photoelectrons being detected for $\lambda_i$ expected number of detected photoelectrons, $P_Q(Q_i|n)$ is the probability of seeing charge $Q_i$ given $n$ photon hits, and $P_T(t_i|n)$ is the probability of observing a PMT trigger at time $t$ given $n$ photon hits. 

The heart of the fitter lies in the first probability term, which is calculated based on Monte Carlo simulations. Muons were simulated at discrete impact parameter values with random directions through the detector. These simulations were used to create lookup tables for how many photoelectrons are expected to be detected by a PMT at a given position with respect to a muon track with a given impact parameter. 

\begin{figure}[tbp]
\includegraphics[width = \figuresize,keepaspectratio=true]{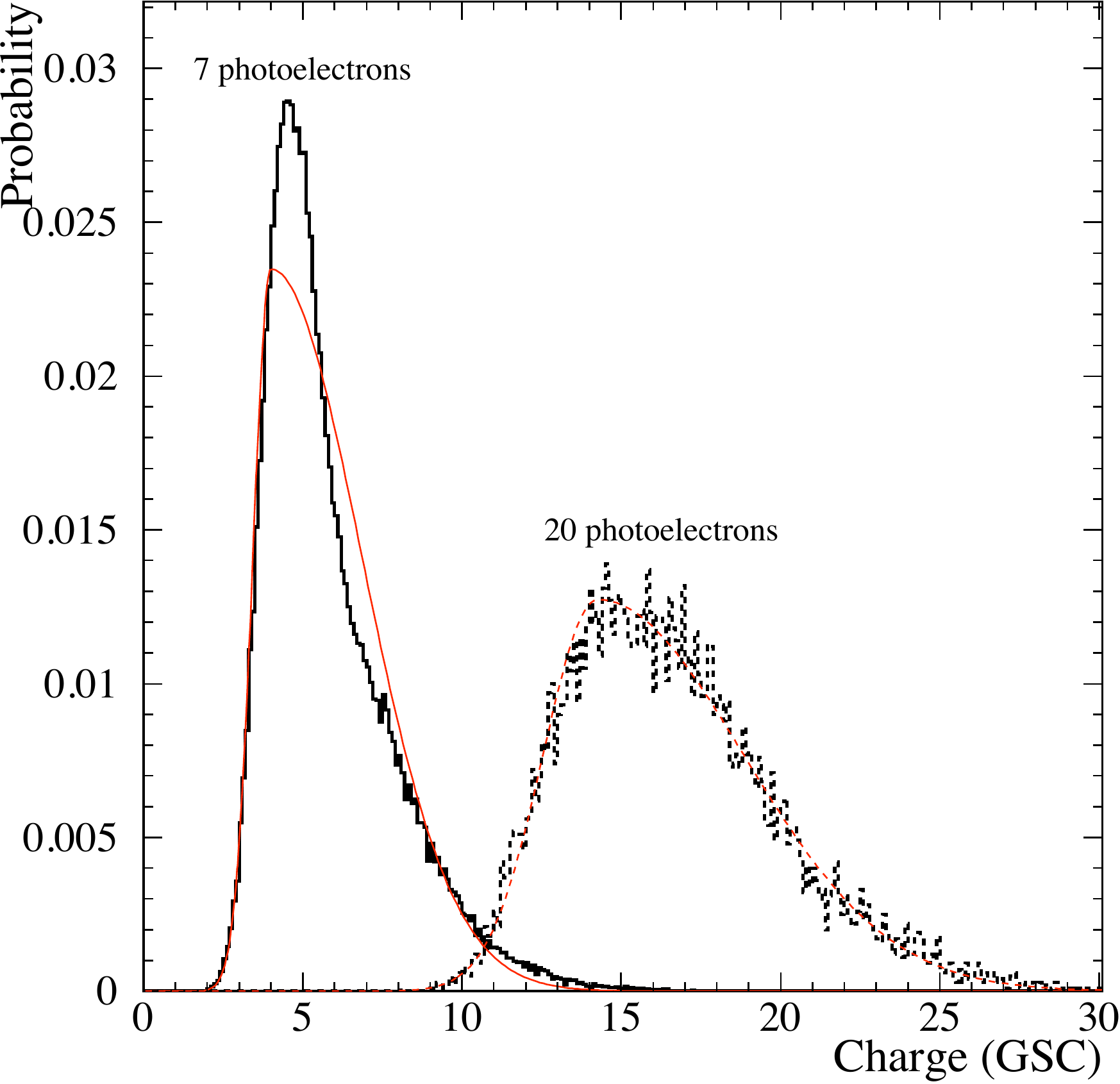}
\caption[FTI Charge Probability Comparison]{\label{fig:Q_prob} The normalized PMT charge distribution measured in (scaled) pedestal-subtracted ADC charge for the case of 7 and 20 photoelectrons striking a single PMT. The red line indicates the prediction from the charge parameterization model used in the reconstruction.}
\end{figure}

The second term further refines the fit by including the charge information from the PMTs, and allows an estimate of the total energy deposited by the muon, correcting for offline PMTs and the neck of the detector. This probability was calculated by simulating multiple photon hits on all of the PMTs in SNO. For a given number of photon hits, the resulting charge distribution is modeled as an asymmetric Gaussian with the widths extracted from simulations.
This fit model agrees well with the simulations for many photon hits, and acceptably for few photon hits (see Figure~\ref{fig:Q_prob}).

The third term in the likelihood refines the fit by including the PMT timing. For each PMT, the time residual can be calculated as:
\begin{equation}
t_{res} = t_{PMT,i} - t_0 - \frac{d_1}{c} - \frac{d_2}{c_D}
\end{equation}
where $t_{PMT,i}$ is the recorded time on a given PMT, $t_0$ is the time offset term in the likelihood fit, $d_1$ is the distance the muon travels within the detector before emitting the Cherenkov photon, $c$ is the speed of light in vacuum, $d_2$ is the distance the Cherenkov photon traveled, and $c_D$ is the average speed of light in D$_2$O/H$_2$O medium (21.8 cm/ns). The Cherenkov photon is assumed to have an angle of $42^\circ$ with respect to the muon track, making $d_1$ and $d_2$ well-defined. The probability of the time residual is modeled as a Gaussian centered at zero with corrections to include estimates of prepulsing and late light as a function of the number of photon hits. 

The SNO muon fitter maximizes the likelihood function for the impact parameter, direction, deposited energy, and timing offset using the method of simulated annealing with downhill simplex~\cite{bib:simplex}. After determining the parameters that maximize the likelihood, a set of data quality measurements are used for background rejection. 

\begin{figure}[tbp]
\includegraphics[viewport=0 0 550 550, width=\figuresize]{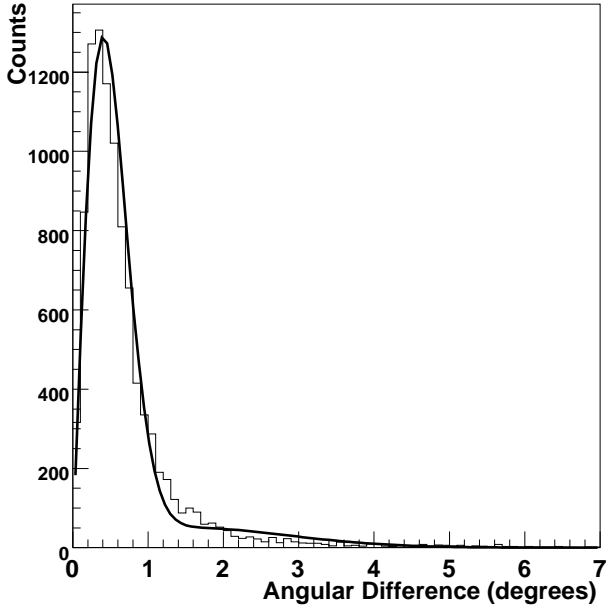}
\caption[Muon Fitter Angular Misreconstruction Simulation] {\label{fig:MCAng} The angular difference (as defined in Eq.~\ref{eq:thetaMR}) of Monte-Carlo muon tracks through the SNO detector (solid histogram). The angular distribution is fit to the function outlined in Eq.~\ref{eq:xtwogaus1} (solid line). The results from the fit are given in Table~\ref{tab:muon_recon}.}
\end{figure}

The muon fitter is found to have good reconstruction accuracy on simulated muons. Figure~\ref{fig:MCAng} shows the angle $(\theta_{mr})$ between the Monte Carlo generated muon direction ($\vec{u}_g$) and the reconstructed muon direction ($\vec{u}_r$): 
\begin{equation}
  \theta_{mr} = \cos^{-1}(\vec{u}_g \cdot \vec{u}_r)
  \label{eq:thetaMR}
\end{equation}
This is fit to a weighted double Gaussian function:
\begin{equation}
p(\theta) = A\theta \left[ f e^{-\frac{\theta^2}{2\sigma^2}}+ (1-f) e^{-\frac{\theta^2}{2(m \sigma)^2}} \right]
\label{eq:xtwogaus1}
\end{equation}

The additional $\theta$-dependence is introduced in order to account for the phase space available. 

\begin{figure}[tbp]
\includegraphics[viewport=0 0 550 550, width=\figuresize]{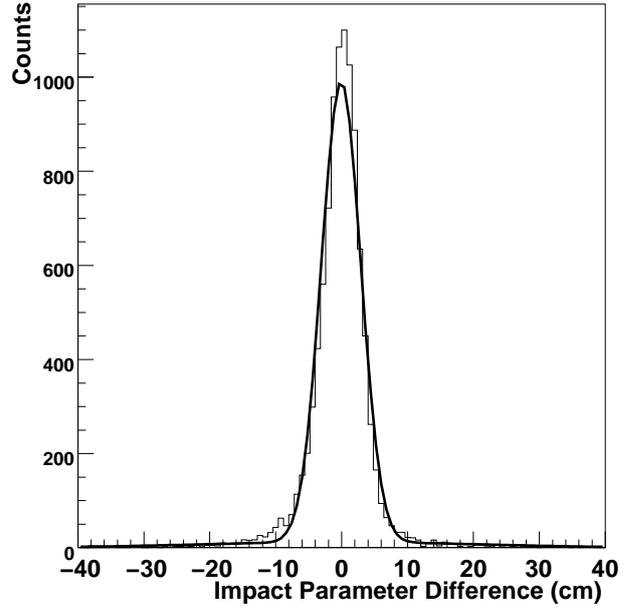}
  \caption[Muon Fitter Impact Parameter Misreconstruction Simulation]{\label{fig:MCImp} The impact parameter difference of Monte-Carlo muon tracks through the SNO detector (solid histogram). The distribution is fit to the function outlined in Eq.~\ref{eq:twogaus1} (solid line). The results from the fit are given in Table~\ref{tab:muon_recon}.}
\end{figure}

The fit parameters are summarized in Table \ref{tab:muon_recon}. Although the tails are non-Gaussian, this fit gives a reasonable estimate for the uncertainty for the angular resolution. Figure~\ref{fig:MCImp} shows the impact parameter reconstruction accuracy. The distribution is fit to the sum of two Gaussians:

\begin{equation}
p(x) = A \left[ f e^{-\frac{(x-\mu)^2}{2\sigma^2}}+(1-f) e^{-\frac{(x-\mu)^2}{2(m\sigma)^2}} \right]
\label{eq:twogaus1}
\end{equation}
with the fit parameters also summarized in Table \ref{tab:muon_recon}. Monte Carlo studies show that the reconstruction accuracy of the muon direction and impact parameter are uncorrelated.

\begin{table}
\centering
\begin{tabular}{|c|c|c|c|c|}
\hline
&$\mu$&$\sigma$&$1-f$&$m\sigma$\\
\hline
\hline
Angular&\multirow{2}{*}{$0^\circ$ (fixed)}&\multirow{2}{*}{$0.4^\circ$}&\multirow{2}{*}{0.01}&\multirow{2}{*}{$1.6^\circ$}\\
 Difference&&&&\\
\hline
Impact Parameter&\multirow{2}{*}{-0.08 cm}&\multirow{2}{*}{3.0 cm}&\multirow{2}{*}{0.012}&\multirow{2}{*}{21 cm}\\
Difference&&&&\\
\hline
\end{tabular}
\caption[Muon Fitter Accuracy from Monte Carlo]{\label{tab:muon_recon} Accuracy of the muon fitter based on Monte Carlo simulations. Fit parameters for mean ($\mu$), widths ($\sigma$ and $m \sigma$), and relative weight ($1-f$) are given in Equations \ref{eq:xtwogaus1} and \ref{eq:twogaus1}.}
\end{table}

\section{The External Muon System}
\label{sec:EMuS}

\begin{figure}[tbp]
  \includegraphics[width=\figuresize]{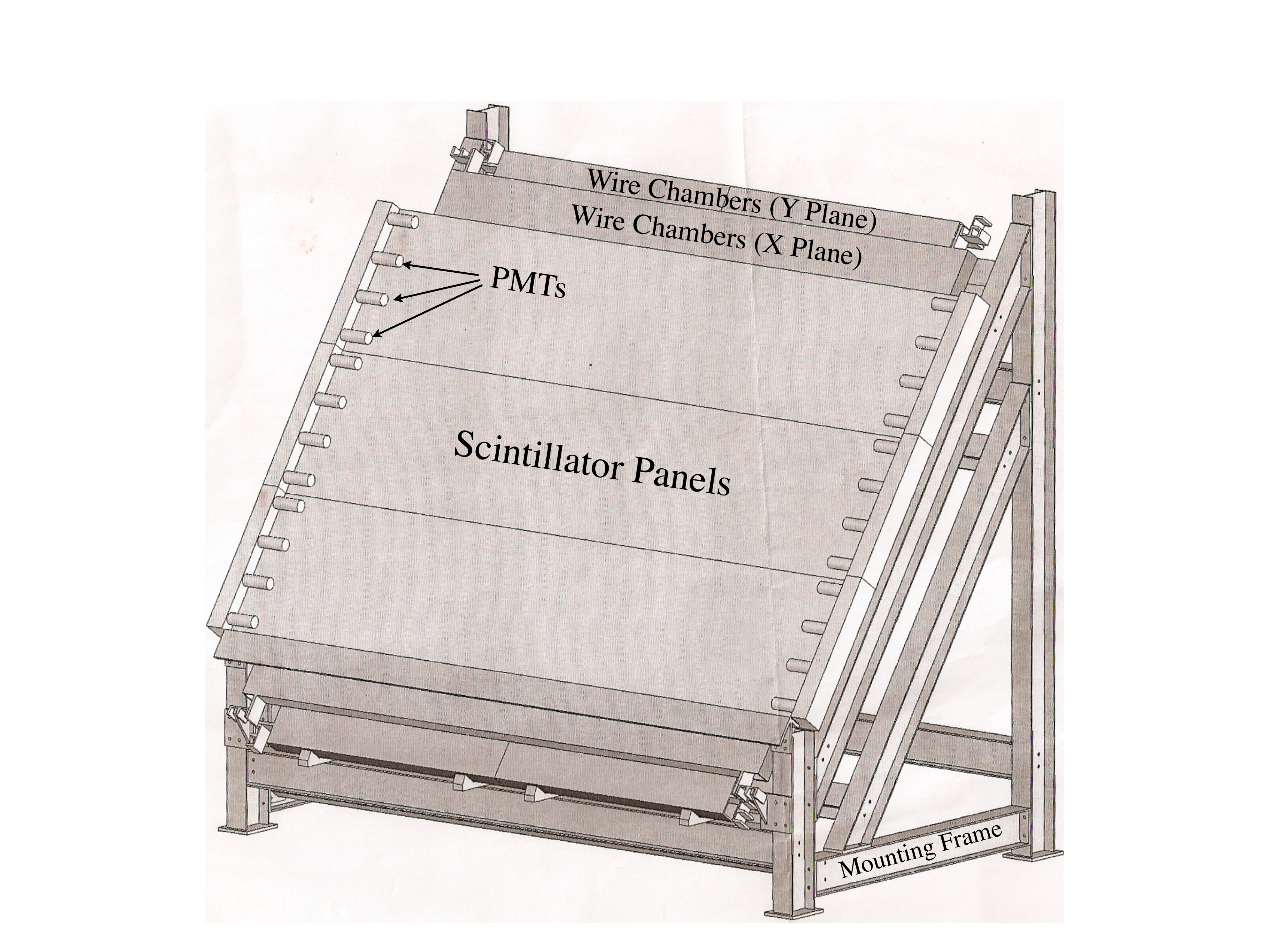}
  \caption{\label{fig:experiment} Diagram of the EMuS detector. See the text for more details.}
\end{figure}
	
\begin{figure}[tbp]
  \includegraphics[width=\figuresize]{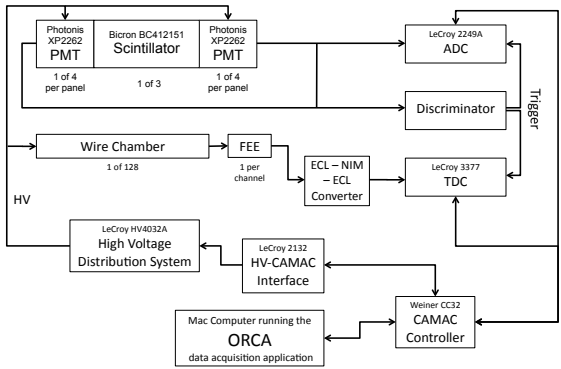}
  \caption{\label{fig:electronics} Diagram of the EMuS electronics system. See the text for more details.}
\end{figure}

The External Muon System consists of a series of 128 single-wire chambers arranged into four planes and triggered by three large scintillator panels (see Figure~\ref{fig:experiment}). The wire chamber cells and electronics were provided by the University of Indiana. Each cell is 7.5~cm wide and has a square cross-section with the corners trimmed into a near-octagonal shape. The cells are 2.564~m in length and possess a single 50~$\mu$m diameter tungsten wire running through the center.  The wire is held at a positive potential of 2500~V (2700~V) while running on the surface (underground) for electron drift and collection. A gas mixture of 90\%Ar-10\%CO$_2$ was used in order to achieve high efficiency and stability, and to meet safety regulations for underground operations.

When a muon passes through the system, it deposits energy in the scintillator and ionizes atoms in each of the wire chambers it passes through. The scintillator converts the energy into light that is then detected by PMTs in a fast process ($\sim$ns). In the wire chambers, the high voltage draws the ionization electrons to the wire in a slow drift process ($\sim\mu$s). The drift time is proportional to the closest distance between the muon track and the wire, allowing track reconstruction using timing and position. The measured drift time for each wire is the time difference between when the scintillator fired and when the drift electrons reached the wire. 

The scintillator consists of three large rectangular panels ($350\times70\times5$ cm$^3$) which cover the active region of the EMuS detector. The panels were acquired from the KARMEN neutrino experiment~\cite{Gemmeke:1990ix}, and consisted of Bicron BC412 scintillator read out at each end by four Photonis~XP2262 PMTs. The signals from the PMTs were sent to a LeCroy~2249A Analog to Digital Converter (ADC) and a discriminator. If both ends of a panel fire in coincidence, a start signal was sent to the wire readout modules, and the ADC modules recorded the pulse-height of each PMT. 

Each wire chamber was monitored by an individual Front-End Electronics (FEE) card which outputted an ECL signal if a pulse is detected on the wire.  The ECL signal was sent to a LeCroy~3377 Time to Digital Converter (TDC) with a readout window of 4.1~$\mu$s. In order to mitigate high levels of electronic noise in the pre-amplifiers, the readout cables were sent through an additional ECL-NIM-ECL converter (see Figure~\ref{fig:electronics}). 

The EMuS system was deployed on the deck of the SNO experiment, 12~m above and 3~m west of the center of the detector. Due to space and solid-angle considerations, the planes were inclined at a 55$^\circ$ from horizontal. A survey was performed to determine the position of each of the wires with respect to the SNO detector.   The dominant sources of uncertainty associated with the wire positions relative to the SNO detector are summarized in Table~\ref{tab:errors}.  The largest uncertainty stems from determining X-Y coordinates of the EMuS detector.  By comparing survey results with other known location markers at the detector, the X-Y coordinate was determined to better than $\pm 0.53$~cm. The reference point used for the Z-coordinate of the detector was only known to $\pm$0.32 cm, and thus added as an uncertainty to the EMuS location. Other uncertainties on the locations of the wires included uncertainties on the floor level, the placement of the wires within the modules, the spacing between wires, and the gaps between the modules. These additional uncertainties do not apply equally to all wires, and have a maximum combined value of $\pm 0.30$~cm. The final uncertainty on the SNO-EMuS coordinate translation based on this survey was $\pm 0.68$ cm.

\begin{table}[b]
\centering
\begin{tabular}{lc}
SNO X-Y Coordinate&0.53 cm\\
SNO Z Coordinate&0.32 cm\\
Floor Level*&0.17 cm\\
Wire Placement&0.08 cm\\
Wire Spacing*&0.18 cm\\
Gaps Between Modules*&0.14 cm\\
Time to Radius Conversion&0.28 cm\\
\hline
Overall&0.74 cm\\
\end{tabular}
\caption{\label{tab:errors}Uncertainties associated with wire positioning. Uncertainties marked by an * do not apply to all wires.}
\end{table}

\subsection{Time to Radius Conversion}

Well-determined models of electron drift and diffusion in a gas \cite{bib:drift} predict that the timing of a wire chamber hit with respect to the scintillator trigger can be used to measure the distance of closest approach of the muon. This time-to-radius conversion function, $r(t)$, has been simulated and measured for the EMuS system. 

\begin{figure}[tbp]
  \includegraphics[viewport=9 7 565 370, width=\figuresize]{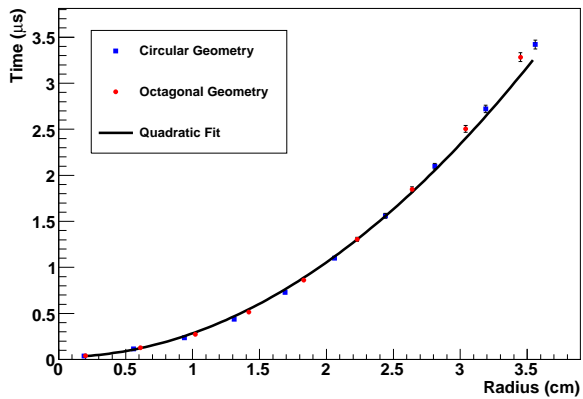}
  \caption[EMuS Simulated Drift Time]{\label{fig:Garfield} The drift time for simulated electrons inside the EMuS wire chambers plotted as a function of starting radius. The plot shows drift times for both circular (boxes) and octagonal (circles) cross-sectional geometries.  The quadratic fit (solid line) is accurate to within 5\% at the maximum simulated radius.}
\end{figure}

The Garfield gas simulation \cite{bib:garfield} was used to generate expected $r(t)$ curves as a function of gas pressure and applied voltage. The code was not able to perfectly model the shape of the wire chambers so two similar geometries were used to check the effects of this imperfect modeling: a circle with radius 3.75 cm, and a regular octagon with a longest radius of 4.06 cm. Simulated electrons were generated at 10 points along the longest radius, and the mean drift time for each point was calculated. Figure~\ref{fig:Garfield} shows that the two $r(t)$ curves agree to within 2\%. A parabolic fit to this data is accurate to 5\%. 

\begin{figure}[tbp]
  \vspace{0.75cm}
  \includegraphics[viewport=9 7 565 530, width=0.70\columnwidth]{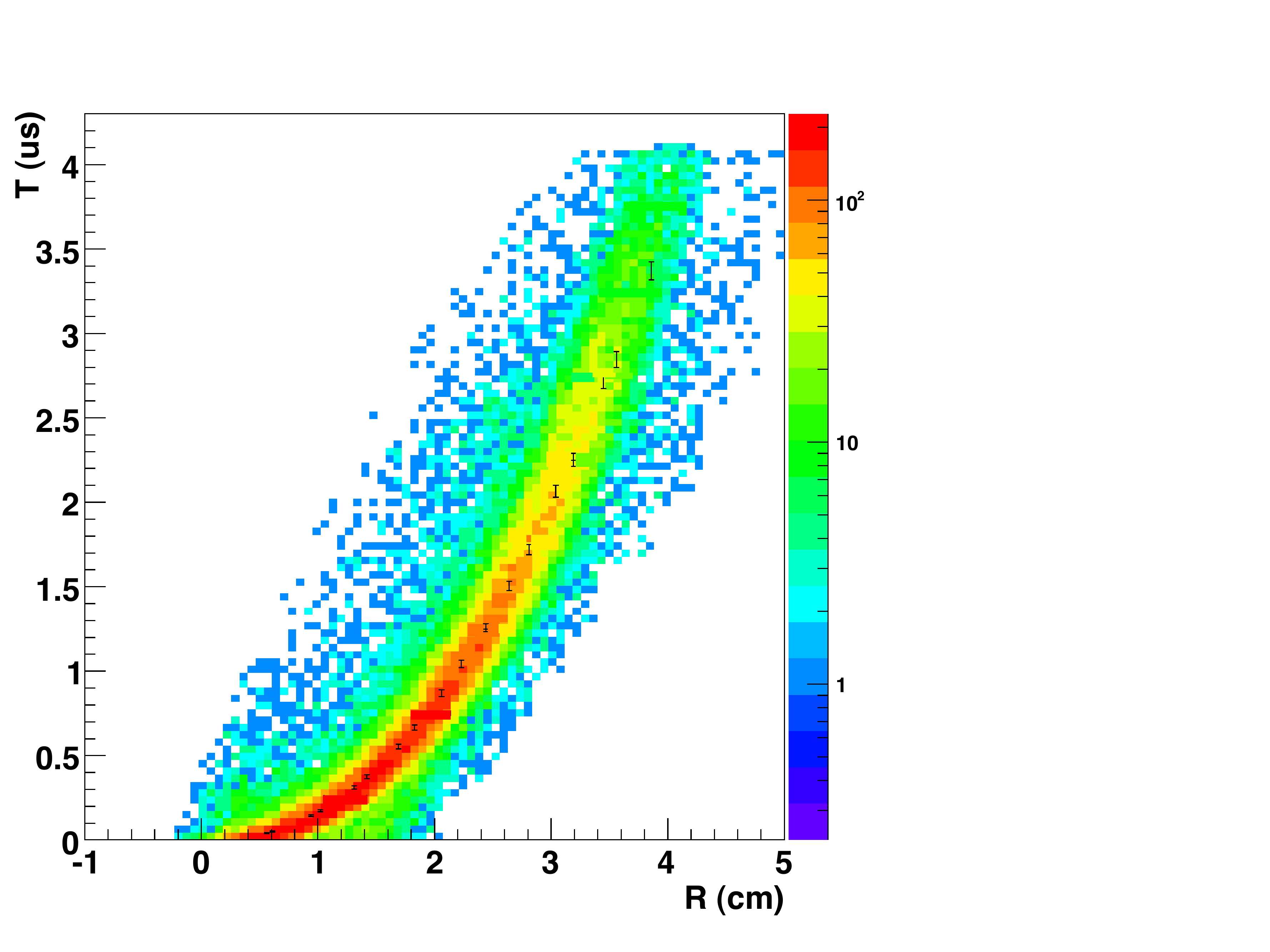}
  \caption[EMuS Drift Time Measurement]{\label{fig:rtdata}Drift time as a function of radius for data taken at Bates Laboratory (surface measurement). The color axis indicates the number of events that reconstruct with the given radius and time. The vertical error bars are Garfield simulations of the drift time.}
\end{figure}

In order to directly measure the $r(t)$ curve, the EMuS system was run on the surface at the MIT-Bates Linear Accelerator Center in Middleton, MA. Candidate muon tracks are selected if they pass through two adjacent chambers on two parallel planes. A series of data cleaning cuts are applied to remove hit pairs created by noise and accidental triggers. Since the positions of the wire chambers that fire are known, an estimate of the angle of the muon trajectory ($\theta$) can be calculated. Once the angle is known, the radii of closest approach are related as: 
\begin{equation}
  R_1+R_2=D\cos{\theta}
  \label{eq:rtheta}
\end{equation}
where $D$ is the distance between each wire.  A trial $r(t)$ function ($\rho(t)=at^2+b$) is used to estimate $R_2$ as a function of the time from the other chamber:
\begin{equation}
  R'_2=D\cos{\theta}-\rho(t_1)
  \label{eq:rtime1}
\end{equation}
A least-squared parameter $B$ is constructed
\begin{equation}
  B=(\rho(t_2)-R'_2)^2
  \label{eq:B}
\end{equation}
\noindent and then minimized. The resulting $r(t)$ curve is shown in Figure~\ref{fig:rtdata}. Slices in time show a Gaussian shape, where the maximum width of these slices is 0.24 cm, which is taken as the uncertainty on the time-to-radius conversion. The fit also extracts a negative time offset of 70 ns, which is caused by delays introduced by the electronic signal chain. This time offset slightly decreases the efficiency for reconstructing events, but does not significantly change the reconstruction accuracy.  Running conditions varied slightly between Bates lab and underground at SNO (mainly due to ambient pressure and operating voltage) and simulations were used to correct for these changes. The extrapolation provides an additional uncertainty of $\pm 0.14$ cm, yielding a total uncertainty of $\pm 0.28$ cm on the time-to-radius conversion model. 

\section{Data Selection}
\label{sec:data_selection}

A number of data quality checks were made to find candidate muons that went through both SNO and the EMuS system.  Six of the EMuS wires were removed from the analysis because of their abnormally low or high trigger rates. A small number of channels had multiple recorded hits in a single event. For such events, only the first hit in time was considered part of the muon track reconstruction algorithm.

EMuS event level cuts were defined to select muon events throughout the run of the experiment. A minimum of three wire planes had to fire in order to ensure proper reconstruction. The event also had to have fewer than 30 wires fired so as to reduce contamination from electrical pickup. Finally, runs with increased human activity above the detector, due to calibrations or source manipulation runs, were removed from the data analysis.  A total of 62 EMuS events passed all run selection criteria. 

To correlate these candidate events with the SNO detector, all of the relevant SNO runs were examined with an event viewer. Of the 62 EMuS events, 32 corresponded to a muon track passing within the volume of the detector confined by the PSUP structure, while 16 corresponded to an event where a muon passed external to SNO's PMT support structure and was therefore seen only by the outward looking PMT tubes. The remaining 14 EMuS events did not traverse the cavity. Of the 32 muon tracks within the SNO detector volume, 30 were properly reconstructed by SNO's muon fitter. The EMuS system ran for 94.6 days of livetime, giving a rate of 0.32 reconstructed coincident events per day. 

\section{EMUs Reconstruction}
\label{sec:recon}

By utilizing tracks that reconstruct in both SNO and the EMuS system, one can determine the final muon track reconstruction accuracy.  A Monte Carlo-based method is used to determine such reconstruction characteristics. For each real data event that is reconstructed in both the SNO and EMuS detector, a series of random test tracks are generated.  These Monte Carlo generated random tracks use the muon track as reconstructed by the SNO detector alone as a seed track, but its vertex and direction are allowed to vary; with up to $\delta \theta \le 10^\circ$ variations in reconstruction angle and up to $\delta b_\mu \le 100$ cm variations in impact parameter.  Subsequently, these generated Monte Carlo tracks are then compared to the hit pattern as recorded in the EMuS tracking chamber. The negative log likelihood value (hereafter referred to as the likelihood) for each generated track is calculated to determine the overall compatibility of the SNO muon reconstruction algorithm with tracks reconstructed in the EMuS system.  The likelihood is given by the following functional form:

\begin{equation}
{\cal L} = \sum_{{\rm wires}~i} \frac{[b_i-\rho(t_i)]^2}{\sigma_i^2}
\label{eq:likely}
\end{equation}

\noindent where $b_i$ is the impact parameter between the simulated track and the $i^{th}$ wire, $\rho(t_i)$ is the expected radius given the TDC time recorded for the wire and $\sigma_i$ is the wire position uncertainty. Wire hits that reconstruct at greater than $5\sigma$ from the main track are essentially removed to avoid reconstruction bias.


\begin{figure}[tbp]
  \includegraphics[viewport=0 0 565 350, width=\figuresize]{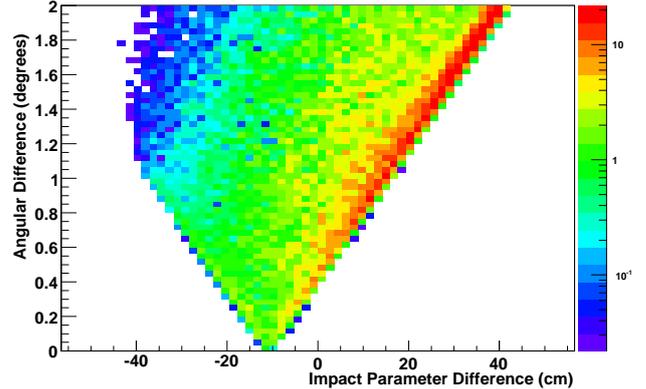}
  \caption[Sample EMuS Track Fit]{\label{fig:emus_lh} Angular difference vs impact parameter difference between SNO's muon fitter and the EMuS system for one event. The color scale indicates the density of possible tracks weighted by their likelihood.}
\end{figure}

\begin{figure*}[tbp]	
\centering
\includegraphics[viewport= 5 300 370 580, width=1.95\columnwidth]{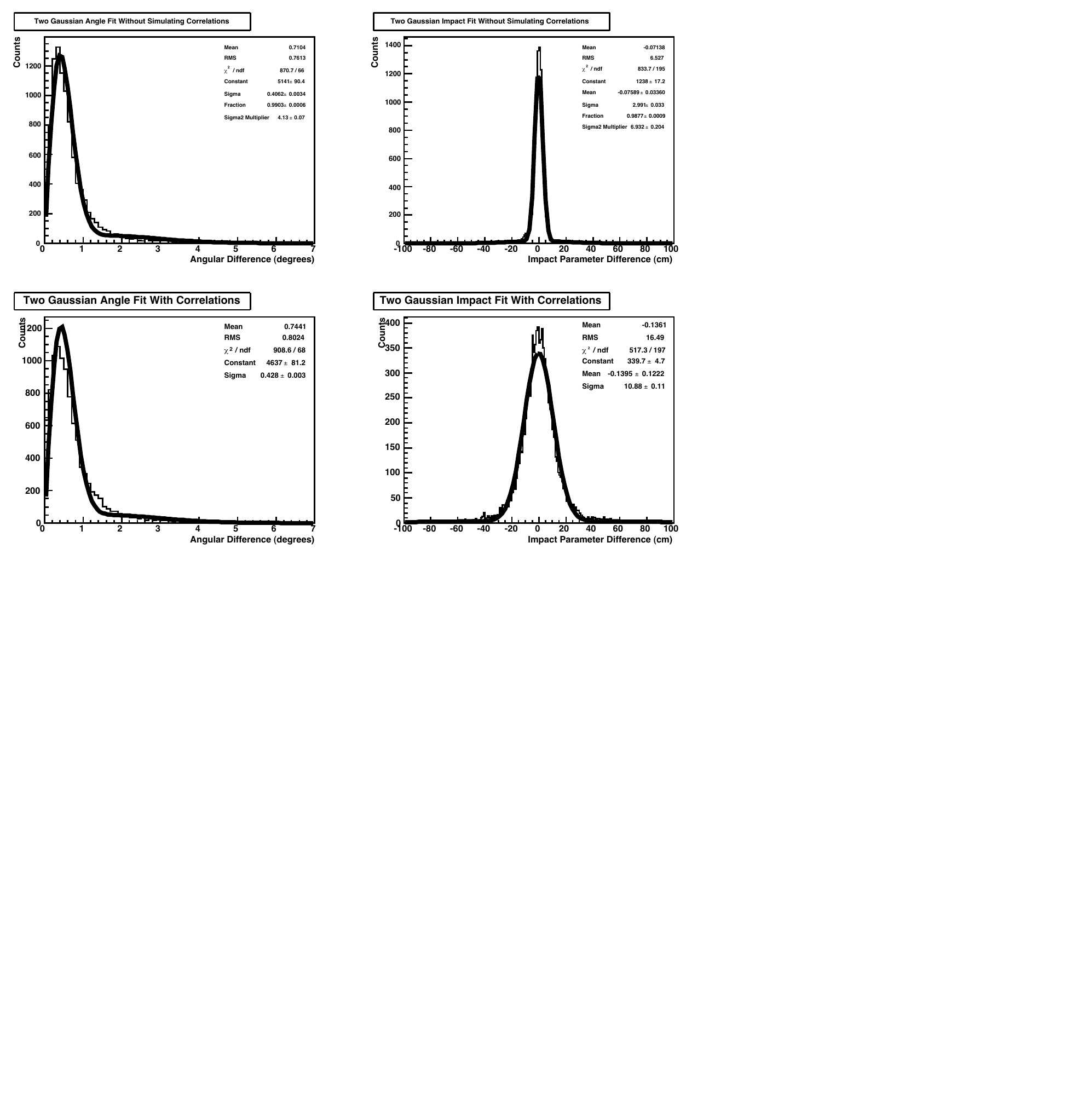} 
\caption[Simulated Misreconstruction Distributions]{\label{fig:Two_Gaus}Results from fitting the angular (left) and impact parameter (right) distributions of the ensemble of the generated simulated data sets according to Equations~\ref{eq:xtwogaus1} and \ref{eq:twogaus1}; respectively. The top plots show the results of fitting the distributions directly from SNOMAN Monte Carlo simulation package without taking into account correlations between angle and impact parameter reconstruction in the EMuS data. The bottom plots show the results with the inclusion of these correlations.}
\end{figure*}

Figure~\ref{fig:emus_lh} shows the most likely tracks for a single event based on this method. The distribution is the projection of a cone, and indicates that there is a degeneracy between the angle and track reconstructed by the EMuS system. This is expected because if the track direction is changed (raising the angular difference) the placement of the track can be changed (raising the impact parameter difference) without significantly altering the hit pattern recorded by the EMuS system. 

Since this ambiguity exists only in the EMuS system and not in SNO's muon tracking algorithm, we can compare tracks reconstructed in the two systems by assuming either (a) the impact parameter is fixed or (b) the reconstructed track direction is fixed.  To test the validity of these assumptions, an ensemble of fake data sets is generated both with and without accounting for track correlations in the EMuS system.  The results from these Monte Carlo tests are shown in Figure~\ref{fig:Two_Gaus}.  Correlations have no effect on the angular mis-reconstruction or the means of the distributions, but they do broaden the impact parameter mis-reconstruction by as much as 10 cm.  We conclude that the EMuS-SNO tracks are sensitive enough to constrain the angular reconstruction and impact parameter bias of the SNO muon fitting algorithm, but not the resolution of the impact parameter reconstruction.

\begin{figure*}[tbp]	
\begin{tabular}{cc}
\includegraphics[viewport= 5 5 560 365, width=\figuresize]{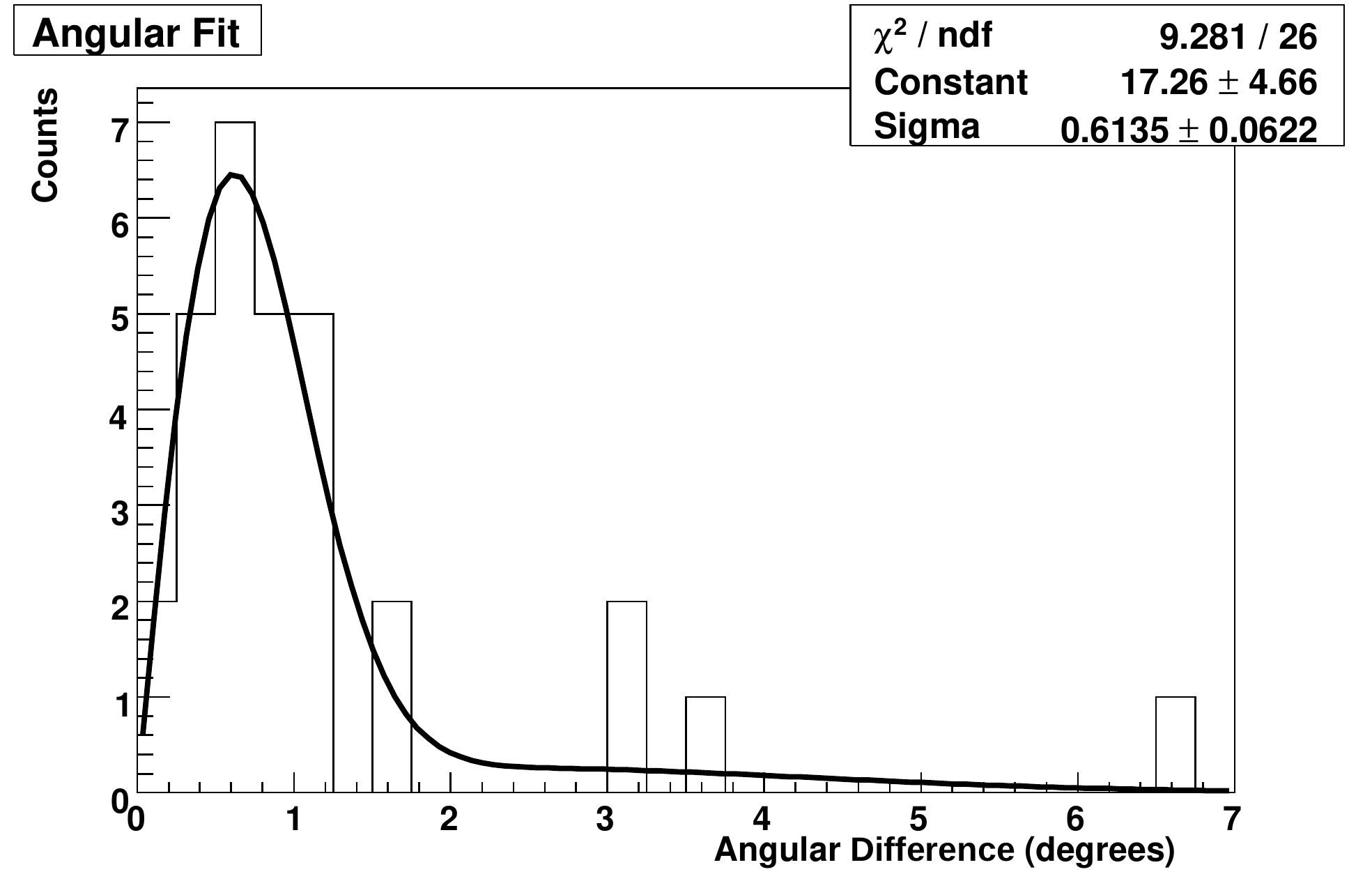} &
\includegraphics[viewport= 5 5 560 365, width=\figuresize]{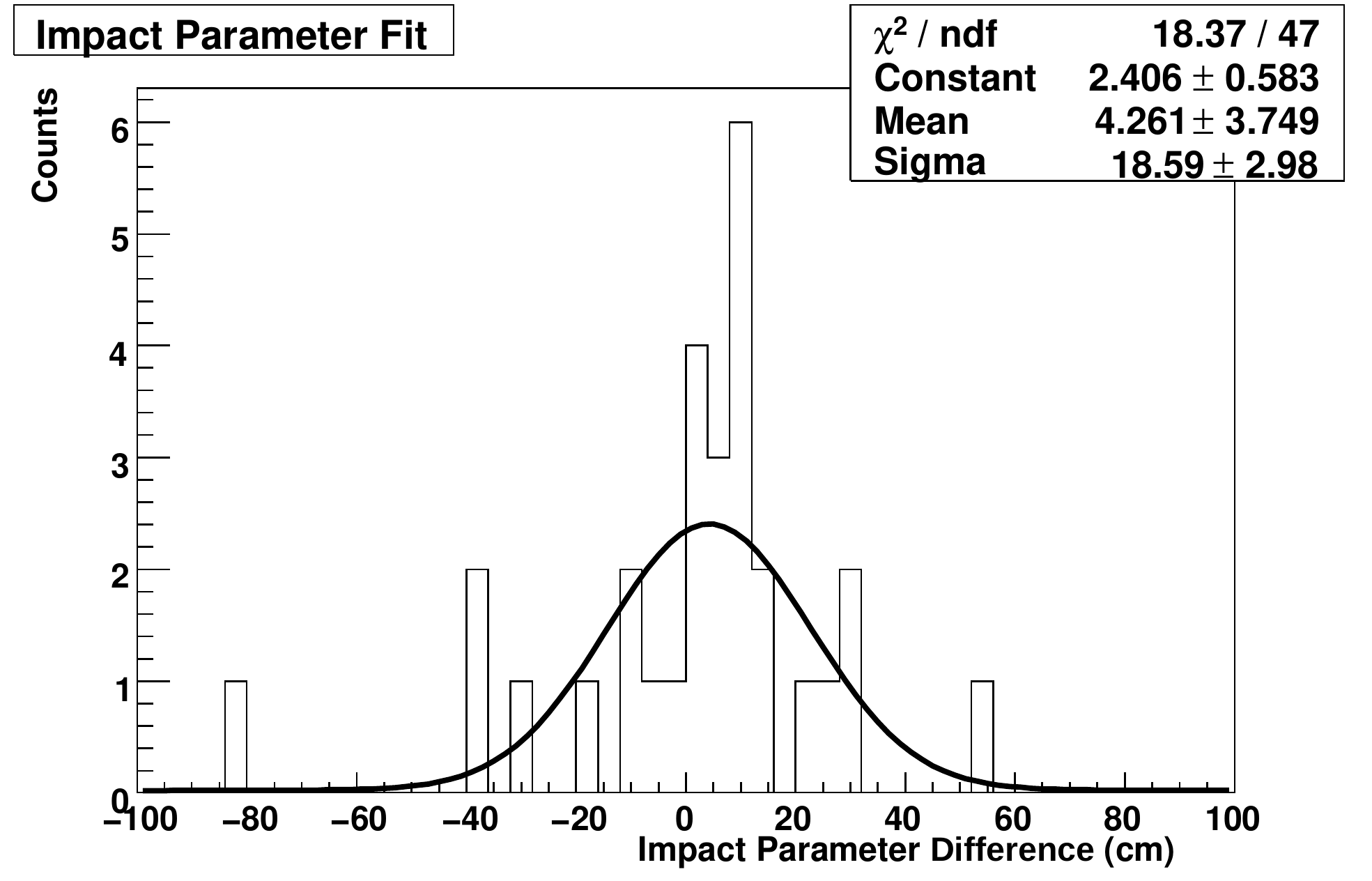} \\
\end{tabular}
\caption[EMuS Impact Parameter Misreconstruction Data]{\label{fig:ang_noshift}Gaussian fit to the data jointly reconstructed by the EMuS-SNO systems.  Figure shows both angular (left) and impact parameter (right) difference.}
\end{figure*}

Figure~\ref{fig:ang_noshift} shows the results of applying the two assumptions to the 30 reconstructed EMuS-SNO events. The data are fitted to the functional forms of Equations~\ref{eq:xtwogaus1}~and~\ref{eq:twogaus1}.  Due to the small number of events, the weights and relative widths of the secondary gaussians are fixed to their values from the earlier simulations. We find that the angular width is $0.61^\circ \pm 0.06^\circ$. The impact parameter bias is $4.2\pm 3.7$ cm, while fit impact parameter width is $18 \pm 11$ cm. 

\section{Conclusions}

The combined data from the SNO detector and the External Muon System have demonstrated that the SNO muon reconstruction algorithm is accurate to the level needed by the neutrino-induced atmospheric flux analysis. The EMuS analysis places a constraint on the angular reconstruction to better than $0.61^\circ \pm 0.06^\circ$ and on the impact parameter bias to better than $4.2\pm 3.7$ cm.  The latter constraint is in good agreement with other methods using cosmic-ray data in SNO~\cite{bib:SNO_atmo}.  We believe the method employed here is a unique, low-cost way to explicitly verify the validity of muon track reconstruction for deep underground experiments.

\section{Acknowledgements}
\label{sec:ack}
This research was supported by: Canada: Natural Sciences and Engineering Research Council, Industry Canada, National Research Council, Northern Ontario Heritage Fund, Atomic Energy of Canada, Ltd., Ontario Power Generation, High Performance Computing Virtual Laboratory, Canada Foundation for Innovation; US: Dept. of Energy, National Energy Research Scientific Computing Center; UK: Science and Technology Facilities Council; Portugal: Funda\c{c}\~{a}o para a Ci\^{e}ncia e a Tecnologia. We would like to thank the University of Indiana, Los Alamos National Laboratory, and K. Eitel for loan of equipment to make the measurement possible.  We would also like to thank the SNO technical staff for their strong contributions and Vale (formerly Inco) for hosting this project.


\begin{thebibliography}{99}
%
\bibitem{bib:SNO_atmo}
B. Aharmim {\em et al.}, Phys. Rev. D 80 (2009) 012001.
%
\bibitem{bib:SK}
Y. Ashie {\em et al.}, Phys. Rev. D 71 (2005) 112005.
%
\bibitem{bib:Minos}
P. Adamson {\em et al.}, Phys. Rev. D 77 (2008) 072002.
%
\bibitem{bib:sno_cal}
B. A. Moffat {\em et al.}, Nucl. Instrum. Meth. A 554 (2005) 255.
%
\bibitem{bib:Doucas}
G.~Doucas {\em et al.} Nucl. Instrum. Methods A 370:579 (1996).
%
\bibitem{bib:sno_nim} 
J.~Boger {\em et al.}, Nucl. Instrum. Meth. A 449 (2000) 172.
%
\bibitem{bib:sno_cal2}
M. R. Dragowsky {\em et al.}, Nucl. Instrum. Meth. A 481 (2002) 284.
%
\bibitem{bib:sno1}
Q.R. Ahmad {\em et al.}, Phys. Rev. Lett. 87 (2001) 071301.
%
\bibitem{bib:sno2}
Q.R. Ahmad {\em et al.}, Phys. Rev. Lett. 89 (2002) 011301.
%
\bibitem{bib:sno3}
Q.R. Ahmad {\em et al.}, Phys. Rev. Lett. 89 (2002) 011302.
%
\bibitem{bib:sno4}
B. Aharmim {\em et al.},  Phys. Rev. C 75 (2007) 045502.
%
\bibitem{bib:sno5}
S.N. Ahmed {\em et al.}, Phys. Rev. Lett. 92 (2004) 181301.
%
\bibitem{bib:sno6}
B. Aharmim {\em et al.}, Phys. Rev. C 72 (2005) 055502.
%
\bibitem{bib:sno7}
B. Aharmim {\em et al.}, Phys. Rev. Lett. 101 (2008) 11130.
%
\bibitem{bib:orca}
M.A. Howe {\em et al.}, IEEE Trans. Nucl. Sci. 51 (2004) 878.
%
\bibitem{bib:simplex}
W. H. Press, S. A. Teukolsky, W. T. Vetterling, and B. P. Flannery, Numerical Recipes 
in Fortran, Cambridge University Press, 2nd ed (1992). 
%
\bibitem{Gemmeke:1990ix}
  H.~Gemmeke {\it et al.},
  Nucl.\ Instrum.\ Meth.\  A  289, 490 (1990).
%
\bibitem{bib:drift}
A. Peisert and F. Sauli, CERN-84-08 (1984).
%
\bibitem{bib:garfield}
R. Veenhof, ``Garfield", CERN Program Library (1998).

\end{thebibliography}
\end{document}